\documentclass[twocolumn,showpacs,preprintnumbers,amsmath,amssymb]{revtex4}

\usepackage{graphicx}
\usepackage{graphicx,epstopdf}
\usepackage{dcolumn}
\usepackage{bm}

\begin{document}


\title{Basin stability for chimera states}
\author{Sarbendu Rakshit$^1$}
 \author{Bidesh K. Bera$^1$}
\author{Matja{\v z} Perc$^{2, 3}$}
\author{Dibakar Ghosh$^1$}
\email{diba.ghosh@gmail.com}
\affiliation{$^1$Physics and Applied Mathematics Unit, Indian Statistical Institute, 203 B. T. Road, Kolkata-700108, India\\
$^2$Faculty of Natural Sciences and Mathematics, University of Maribor, Koro{\v s}ka cesta 160, SI-2000 Maribor, Slovenia\\
$^3$CAMTP -- Center for Applied Mathematics and Theoretical Physics, University of Maribor, Krekova 2, SI-2000 Maribor, Slovenia}

\date{\today}

\begin{abstract} Chimera states, namely complex spatiotemporal patterns that consist of coexisting domains of spatially coherent and incoherent dynamics, are investigated in a network of coupled identical oscillators. These intriguing spatiotemporal patterns were first reported in nonlocally coupled phase oscillators, and it was shown that such mixed type behavior occurs only for specific initial conditions in nonlocally and globally coupled networks. The influence of initial conditions on chimera states has remained a fundamental problem since their discovery. In this report, we investigate the robustness of chimera states together with incoherent and coherent states in dependence on the initial conditions. For this, we use the basin stability method which is related to the volume of the basin of attraction, and we consider nonlocally and globally coupled time-delayed Mackey-Glass oscillators as example. Previously, it was shown that the existence of chimera states can be characterized by  mean phase velocity and a statistical measure, such as the strength of incoherence, by using well prepared initial conditions. Here we show further how the coexistence of different dynamical states can be identified and quantified by means of the basin stability measure over a wide range of the parameter space.        
\end{abstract}


\maketitle


\section*{Introduction}
During the last decade, one of the most interesting research areas concerning identically coupled oscillators is the coexistence of coherent and incoherent states. This interesting spatiotemporal behavior was first observed by Kuramoto and Bottogtokh \cite{kuramoto} in a network of nonlocally coupled Ginzburg-Landau oscillators with exponential coupling functions. Later, Abrams and Strogatz named this intriguing spatiotemporal state the chimera state \cite{strogatz}. Initially, it was believed that a nonlocal coupling topology is the necessary requirement for the existence of chimera states in complex networks. Subsequently, however, it has been shown that this condition is not absolutely necessary, and that chimera states can also observed in all-to-all \cite{global1,global2,global3,global4,global5,global6,global7} and nearest-neighbor \cite{laing,hr_bera1,hr_bera2,local1} coupled oscillators, even using one-sided local coupling \cite{hr_bera3}. Chimera states were first detected in phase oscillators, after which they were also reported in limit-cycle oscillators \cite{limit,hr_bera2}, chaotic oscillators \cite{chaotic}, chaotic maps \cite{chaotic_map}, hyper chaotic time delay systems \cite{lakshman_measure}, and even in neuronal systems that exhibit bursting dynamics \cite{hr_ijbc,hr_bera1,chimera_modular,hr_bera3}. Most recently, chimera states were also observed in multiplex network \cite{chimera_multiplex,multiplex2,multiplex3,multiplex1}. Depending on the type of symmetry breaking in coupled networks, chimera states can be classified into various categories, such as  amplitude-mediated chimeras \cite{amc}, globally clustered chimeras \cite{gcc}, amplitude chimeras, and chimera deaths \cite{cd_prl}. Moreover, based on the spatiotemporal behavior of coherent and incoherent motion, new terms have been coined such as breathing chimeras \cite{breath1}, imperfect chimeras \cite{imperfect_chi}, traveling chimeras \cite{travelling_chi}, imperfect traveling chimeras \cite{hr_bera3}, as well as spiral wave chimeras \cite{chaotic,spiral_chi}. Apart from the above theoretical research, the robustness of chimera states has also been verified in experiments. The first experimental observations of chimera states have been in optical \cite{opto-elect} and chemical \cite{chemical_exp,chemical_exp2} systems. After that, chimera states have also been observed experimentally in several other systems, such as in electronic circuits \cite{electronic2,electronic}, electrochemical \cite{electro,electro2} and opto-electronic systems \cite{opto_electronic}, boolean networks \cite{boolean}, optical combs \cite{opt_comb}, as well as in mechanical systems \cite{mechanical,imperfect_chi}. Chimera states also occur in real world systems \cite{chimera_rev}, such as in power grids \cite{power1,power2}, social networks \cite{social}, as they have also been observed in unihemispheric slow-wave sleep of some migratory birds and aquatic mammals \cite{uhsws1,uhsws2}. Related to the latter, during slow-wave sleep one half of the brain rests and the neuronal oscillators are synchronized, while the other half is awake and the neurons are thus desynchronized.

\par Ever since its discovery, the initial conditions have played a crucial role in the existence of chimera states in complex dynamical networks. In many systems, chimeras and fully synchronized states coexist, but chimera states emerge only for specific initial conditions, and they do not appear via spontaneous symmetry breaking. A proper choice of initial conditions is thus required for chimera states in complex dynamical networks \cite{ba_njp}. However, it has also been shown that chimeras can emerge from random \cite{random} or quasi-random \cite{random2} initial conditions. For nonlocal and local (nearest neighbor) interactions, each node in the coupled network is not interacting to all the other nodes at a time, so there is a chance for chimera sates appearing with a proper choice of initial conditions. Conversely, for globally (all-to-all) coupled networks, there is a comparatively low chance of symmetry breaking if using simple scalar diffusive coupling. But there is a possibility of chimera states appearing in globally coupled networks using diffusive interactions if the system exhibit multi-stable behavior. Recently, in \cite{global2}, it was shown that chimera states in globally coupled networks can nevertheless emerge by means of the so-called intensity induced mechanism. Based on the above considerations, one can conclude that initial conditions in a network of coupled oscillators systems are crucial for the emergence of chimera states. However, the probability of the emergence of chimera states together with coherent and incoherent states at a particular coupling strength for different initial conditions has not yet been studied and deserves special attention.

\par Using this as motivation, we study how robust different chimera states are with respect to the initial conditions. To that effect, we adopt the concept of the basin stability (BS) \cite{bs_nature}, which is closely connected to the volume of the basin of attraction. The BS approach is nonlinear and nonlocal, but easily applicable even to higher dimensional systems. Recently, BS has been used to quantify different stable steady states in non-delayed \cite{our2017} and time-delay systems \cite{bs_timedelay}, as well as in power grid systems \cite{pwr_grd1,pwr_grd2} and various other fields of science \cite{njp,epjst,bspre}. In this work, we characterize the incoherent, chimera and coherent states by calculating the strength of incoherence and the time average mean phase velocity profile. Based on the value of the strength of incoherence, we develop the basin stability measure for different dynamical states.  We consider time delayed Mackey-Glass systems \cite{mg} to explore such phenomena using nonlocal and global interactions. We numerically investigate how different dynamical states coexist at a fixed value of  coupling strength and for different initial history functions. The variation of basin stability due to different coupling strengths is investigated in detail. We conclude with a discussion of how our method is applicable to quantify different dynamical states in coupled oscillators, also for other types of coupling configurations.

\section*{Results}
The following sections are devoted to the basin stability measure for different dynamical states such as incoherent, coherent and chimera states under two coupling configurations, namely nonlocal and global. Using nonlocal coupling, our main emphasis will be to identify the variation of basin stability in the parameter region of coupling strength $\epsilon$ and coupling radius $R$. Later, chimera states emerge in globally coupled oscillators induced by intensity mechanism recently proposed by Chandrasekhar et al. \cite{global2} and we will discuss the variation of basin stability by changing the coupling strength. We will explore both the phenomena in a network of coupled Mackey-Glass systems.

\subsection*{Nonlocally Coupled Network}

In order to exemplify the basin stability measure, we first consider a system of time-delayed Mackey-Glass systems coupled through nonlocal fashion with finite coupling radius. The mathematical models are given by
\begin{equation}
\begin{array}{lcl}
\dot x_i=-ax_i+\frac{bx_i(t-\tau)}{1+x_i^{10}(t-\tau)}+\frac{\epsilon}{2p}\sum\limits_{j=i-p}^{j=i+p}(x_j-x_i), \; i=1,...,N,
\end{array}
\end{equation}
where $\epsilon$ is the coupling strength, $N$ is the total number of oscillators in the network, $p$ is the number of nearest neighbor oscillators in each side on a ring coupled with the $i$-th oscillator, $R=\frac{p}{N}$ is defined as the coupling radius. Without coupling (i.e. $\epsilon=0.0$), an isolated oscillator exhibits chaotic behavior for the set of parameters $a=1, b=2$ and $\tau=2$. To simulate the equations of coupled network ($N=100$), we choose `V' shaped constant initial history function of each oscillator, i.e. the history in the interval $[-\tau, 0]$ for $x_i$, as follows $x_{i0}=c(\frac{N}{2}-i)$ for $i=1,2,...,\frac{N}{2}$ and $x_{i0}=c(i-\frac{N}{2})$ for $i=\frac{N}{2}+1,...,N$ where $c$ is a constant. We can change the initial condition by varying the value of $c$. In general, the transition occurs from incoherent to coherent state through a chimera or multi-chimera state in a coupled network with changing the coupling strength $\epsilon$. But by proper choice of $c$, the different dynamical states may coexist in the network (1) for a certain value of $\epsilon$.


\begin{figure}[ht]
		\centerline
			{\includegraphics[scale=0.44]{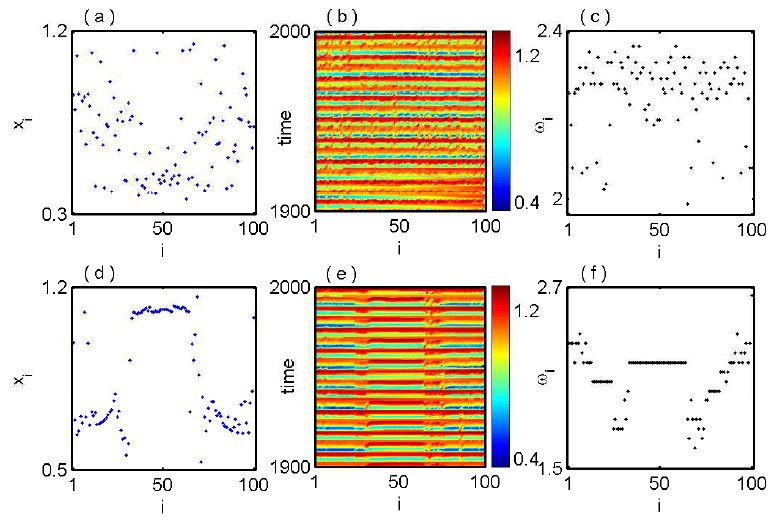}}
	\caption{ Nonlocally coupled Mackey-Glass systems: coexistence of incoherent and chimera states at fixed coupling strength $\epsilon=0.35$ and coupling radius $R=0.3$ for different initial conditions (a, b, c) $c=0.001$ and (d, e, f) $c=0.03$. Left columns show snapshot of amplitude $x_i$ (i=1,2,...,N) in blue points, middle columns for space-time plot and right columns for time average phase velocities $\omega_i$ (black dotted points). }
	\label{fig1}
\end{figure}

\par We fix the value of coupling radius $R=0.3$, coupling strength $\epsilon=0.35$ and vary the constant $c$ in the initial history function. For an exemplary value of $c=0.001$ i.e. for a particular fixed initial condition, the network exhibits an incoherent state and the snapshot of $x_i$ are shown in Fig.~~\ref{fig1}(a).  The values of $x_i$ are randomly distributed in [0.3, 1.2] which shows the incoherent state. Figure~~\ref{fig1}(b) shows the corresponding space-time plot of incoherent state. We also confirm the incoherent state by calculating the time-averaged phase velocities $\omega_i$ (refer to the Method section) in Fig.~~\ref{fig1}(c). The time-averaged phase velocities of all the oscillators are randomly distributed, which indicate the incoherent state for that particular initial condition.  Next we search for another initial condition by changing the  constant $c=0.03$ in the initial history function $x_{i0}$, the snapshot and space-time plot of $x_i$ of all the oscillators in the network (1) are shown in Figs. ~~\ref{fig1}(d) and \ref{fig1}(e) respectively. From these figures it is noted that the whole network (1) breaks into two groups, one coherent and another incoherent group, which re-ensemble a chimera state.  The time average phase velocities of each oscillator in the network are shown in Fig.~~\ref{fig1}(f) and from this phase velocity profile clearly signify the emergence of chimera state.  So, depending on the initial condition of each oscillator, coexistence of incoherent  and chimera state, is observed at a fixed value of coupling strength.
\par
\begin{figure}[ht]
		\centerline
			{\includegraphics[scale=0.44]{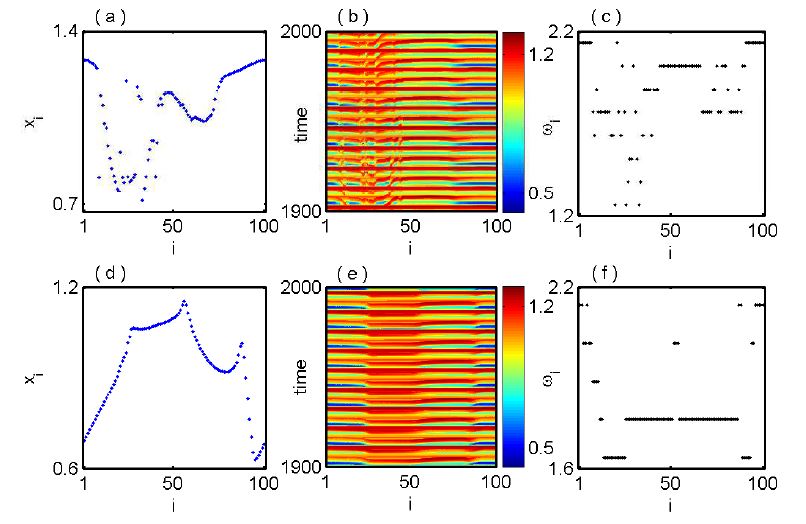}}
	\caption{ Nonlocally coupled Mackey-Glass systems: coexistence of chimera and coherent states at $\epsilon=0.56$ for (a, b, c) $c=0.09$ and (d, e, f) $c=0.08$ respectively. The left, middle and right column panels are respectively show the snapshot of amplitude $x_i$, spatiotemporal plot and mean phase velocity.}
	\label{fig2}
\end{figure}

 Next, we find the similar coexistence of two different dynamical behaviors such as coherent and chimera state at $\epsilon=0.56$.
Figure \ref{fig2}(a) shows the snapshot of amplitude of chimera states for $c=0.09$ and  the corresponding spatiotemporal behavior are shown in Fig.~~\ref{fig2}(b). For the small deviation of initial condition as $c=0.08$, we find a smooth profile of the state variables $x_i$ display in Fig.~~\ref{fig2}(d) and corresponding spatiotemporal plot in Fig.~~\ref{fig2}(e). The mean phase velocity profiles for $c=0.09$ and $c=0.08$ are respectively shown in Figs.~~\ref{fig2}(c) and \ref{fig2}(f),  which are the clear indication of chimera and coherent states.

\par
\begin{figure}[ht]
 		\centerline
 		{	\includegraphics[scale=0.44]{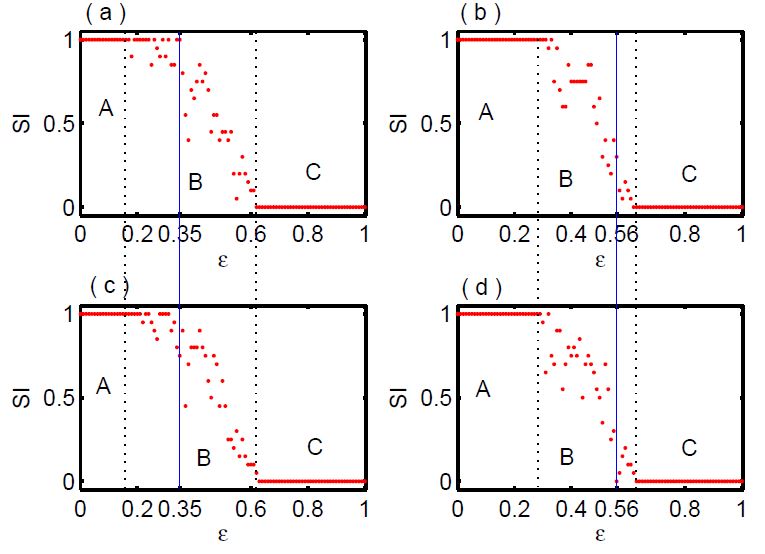}}
 	\caption{ Strength of incoherence is plotted with respect to the coupling strength $\epsilon$ for fixed coupling radius $R=0.3$ and $N=100$. (a) and (c) show the variation of strength of incoherence for different initial conditions at $c=0.001$ and $c=0.03$ corresponding to the spatiotemporal scenarios in Figs.\ref{fig1}(a, b, c) and \ref{fig1}(d, e, f)  respectively. A solid blue line is drawn through the coupling strength $\epsilon=0.35$, along the blue line SI takes the values `1' in (a) where in (c) the values of SI lies between `0' and `1'. In (b) and (d) strength of incoherences are calculated by varying the interaction strength $\epsilon$ using the initial condition of Fig. \ref{fig2}(a, b, c)  ($c=0.09$) and Fig. \ref{fig2}(d, e, f) ($c=0.08$). At the coupling strength $\epsilon=0.56$, a solid blue line is marked and corresponding this particular coupling strength $SI \in (0,1)$ in (b) and SI takes the values `0' in (d). The regions A, B and C are marked as the regions of incoherent, chimera and coherent states respectively.}
 	\label{fig3}
 \end{figure}

 To clearly distinguish of the various dynamical states, we calculate the strength of incoherence (SI) (refer to the Method Section) which is a statistical measure based on the time series of the networks. In this SI measure, SI takes the values 1 and 0 for incoherent and coherent states respectively while SI$\in(0,1)$ for the chimera states. The coexistence of chimera with incoherent and chimera with coherent states are respectively shown in Figs.~~\ref{fig1} and \ref{fig2}. Here we characterize the different dynamical states in Figs. \ref{fig1} and \ref{fig2} for the fixed initial conditions using the strength of incoherence measure and the results are illustrated in Figs. \ref{fig3}(a, c) and \ref{fig3}(b, d) respectively. In Figs. \ref{fig3}(a) and \ref{fig3}(c) strength of incoherence is plotted against the coupling strength $\epsilon$ by taking the initial conditions which are used in Figs. \ref{fig1}(a) (for $c=0.001$) and \ref{fig1}(d) (for $c=0.03$). We draw a solid blue line along the coupling strength $\epsilon=0.35$ in Figs. \ref{fig3}(a) and \ref{fig3}(c) where incoherent and chimera states are coexist. In Fig. \ref{fig3}(a), blue line intersect the SI value at 1 which characterize the incoherent state corresponds to Fig. \ref{fig1}(a), while blue line cuts the SI value in $(0,1)$ in Fig. \ref{fig3}(c) which signifies the chimera state corresponding to the snapshot of Fig. \ref{fig1}(d). Similarly, taking the same initial conditions of Figs. \ref{fig2}(a) and \ref{fig2}(d), we plot the SI with respect to the coupling strength $\epsilon$ in Figs. \ref{fig3}(b) and \ref{fig3}(d). At the coupling strength $\epsilon=0.56$, the value of SI lies in (0,1)  in Fig. \ref{fig3}(b) and cuts the SI line (blue line) at 0 in Fig. \ref{fig3}(d) which are the clear indications of  chimera and coherent states, depicted in Figs. \ref{fig2}(a) and \ref{fig2}(d) respectively. It is noted that the spatial position of coherent and incoherent domain is not fixed and they are highly dependent on initial conditions. The three different regions A, B, and C are the range of coupling strength $\epsilon$ for incoherent, chimera and coherent states respectively in Fig. \ref{fig3}. If we take large number of different initial conditions from the basin of volume then the three different regions A, B and C of coupling strength  may be varied according to the initial condition profiles.  Next our aim is to quantify the above three dynamical states in probabilistic sense by taking different initial states in the phase space volume with the variation of the coupling strength $\epsilon$.
 
\par
\begin{figure}[ht]
	\centerline
		{	\includegraphics[scale=0.31]{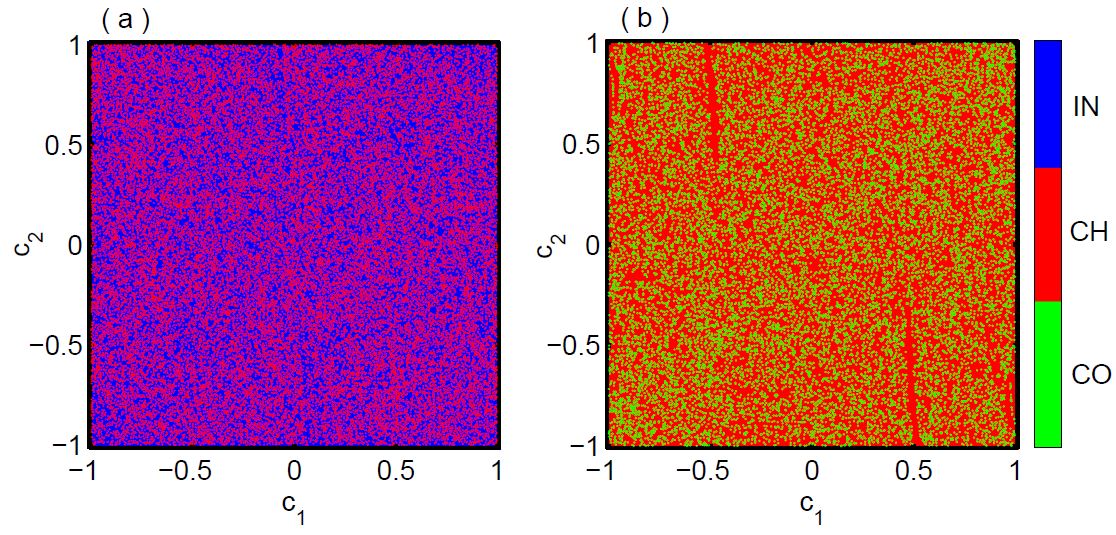}}
	\caption{ Basin of attractions for nonlocally coupled Mackey-Glass systems: (a) coexistence of incoherent and chimera states, $\epsilon=0.33$, (b) coexistence of chimera and coherent states for $\epsilon=0.63$. Here IN, CH and CO stands for incoherent, chimera and coherent states respectively.}
	\label{fig4a}
\end{figure}

 From the above analysis, we conclude that the coexistence of different dynamical states in the network (1) are emerged at a particular value of coupling strength $\epsilon$ with the well prepared initial states. Now it is very interesting to track the basin of attraction of several coexisting states of the coupled network (1). The basin of attraction of time-delayed systems is a function space which has dimension of infinity. So using Hausdroff measure theory, it is possible to evaluate the basin of attraction for time-delayed systems \cite{bs_timedelay}. Such measure is geometrically less intuitive with the traditional concept of basin of attraction of ordinary differential equations \cite{rev}.  To visualize the basin of attractions of the different collective states at a fixed value of $\epsilon$, we project the infinite dimensional initial state space of $i$-th oscillator to a finite dimensional  space [refer to Method section for more details]. We consider the initial history function as $x_i(t)=\phi_i(t)=c_i\sum\limits_{j=0}^{n}(-1)^j\frac{t^j}{j!}$ for $t \in [-\tau, 0]$ where  $n$ is the number of the basis functions and sufficiently large, $c_i \in \mathbb{R}$ and $i=1,2,...,N$. For $N=100$ number of oscillators in the network, we choose $100$ different random values of $c_i$ from the prescribe range. Still it is very complicated to visualize the basin of attractions for all the different values of $c_i$, to avoid this difficulty, we assume the history function of the $i$-th oscillator which spanned by the first two basis constants, i.e., $c_i=\frac{(i-1)c_1+c_2}{i}, i\geq3$ where  $c_1$ and $c_2$ are random numbers in the range $[-1,1]$ so that $x_i(t) \in [-3, 3]$ for $t \in [-\tau, 0]$. In Fig. \ref{fig4a}, we illustrate the basin of attractions for the coexistence of different states in nonlocally coupled Mackey-Glass systems (1) at a fix value of $\epsilon.$  Figures~\ref{fig4a}(a) and \ref{fig4a}(b) show the basin of attractions for the coexistence of incoherent-chimera and chimera-coherent states respectively at the fixed coupling values $\epsilon=0.33$ and $\epsilon=0.63$.  The blue, red and green points are respectively represent the incoherent, chimera and coherent states. The basin of attraction of chimera states are strongly intertwined with incoherent and coherent states which are clearly shown in Figs. \ref{fig4a}(a) and \ref{fig4a}(b).

\par
\begin{figure}[ht]
		\centerline
		{	\includegraphics[scale=0.5]{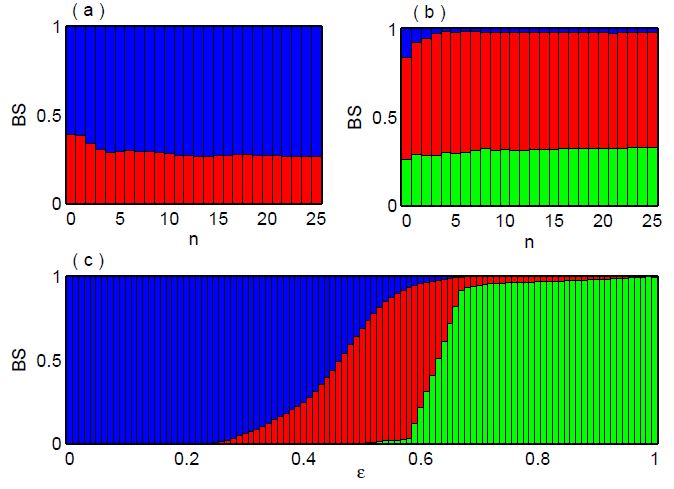}}
	\caption{ Basin stability analysis for nonlocally coupled Mackey-Glass systems: (a) and (b) show the fluctuation of basin stability with respect to the degree of polynomial basis for $\epsilon=0.4$ and $\epsilon=0.625$. (c) The variation of BS of incoherent, chimera and coherent states with respect to the coupling strength $\epsilon$. 	Blue, red and green colors respectively represent the incoherent, chimera and coherent states. }
	\label{fig4}
\end{figure}

 Next, we calculate the basin stability (describe in Method section) for the different dynamical states by varying the coupling strength $\epsilon$.  The basin stability gives us an interesting detail informations to know how stable the different dynamical states such as incoherent, chimera and coherent states against multifarious initial conditions.  Figure \ref{fig4}(a) illustrates the fluctuations of  the basin stability for incoherent and chimera states at $\epsilon=0.4$ by increasing the degree of polynomial basis. From this figure it is show that the fluctuations of BS for incoherent and chimera states are almost saturated for $n\ge 22.$ Next, we check the role of highest degree $n$ in the polynomial basis for another value of coupling strength $\epsilon.$ For this we choose the value of $\epsilon=0.625$ where three states i.e. incoherent, chimera and coherent states are coexist. Figure \ref{fig4}(b) displays the variation of BS for incoherent, chimera and coherent states at $\epsilon=0.625$ represented by blue, red and green colors with respect to the highest degree $n$ of the polynomial basis. From these two figures (Figs. \ref{fig4}(a, b)), it is seen that the fluctuation of BS of their respective dynamical states is tend to be stable value for sufficiently large $n$.  Hence for our coupled system (1) it is enough to take $n=25$ as the highest degree of the initial history function. In Fig. \ref{fig4}(c),  the variation of BS of the above three mentioned dynamical states is delineated with respect to the different values of coupling strength $\epsilon$. For very smaller values of the coupling strength $\epsilon$, the incoherent states is solely dominated showing by blue color and it persists up to $\epsilon=0.2$. In this range of $\epsilon$, there is a high possibility to remember the memory where the final states to be incoherent. After this critical value of coupling parameter $\epsilon \ge 0.2$, the BS of chimera states (represent by red) are appear in small range and gradually increases while incoherent states tend to diminish.  So  annihilation of BS of some state represents a clear signature of the transition point of some states. At $\epsilon\simeq0.5$, the BS of coherent state takes place in the scenario, which means that at this point the possibility of remember of the memory of the incoherent state is very low. For further increasing of the coupling values $\epsilon$, three states are coexists around $\epsilon=0.6$ where incoherent and coherent states emerge tiny portion in the basin volume compare to chimera states. The increment values of $\epsilon$ from $\epsilon\simeq0.63$, the values of BS for chimera state are getting smaller and smaller and there is a slight abrupt transition from chimera to coherent states (marked by green color). Also for more increasing values of $\epsilon \ge 0.63$, the value of BS for incoherent state is completely eliminated and the chimera state appears very small region in the basin volume. Finally for higher values of $\epsilon$, the value of BS for coherent states tend to unit value and acquire more and more space in the basin volume. Such changes of BS measures give an idea of the annihilation of incoherent and chimera states and development of coherent states by changing the coupling strength $\epsilon$.

\par
\begin{figure}[ht]
 		\centerline
 		{	\includegraphics[scale=0.33]{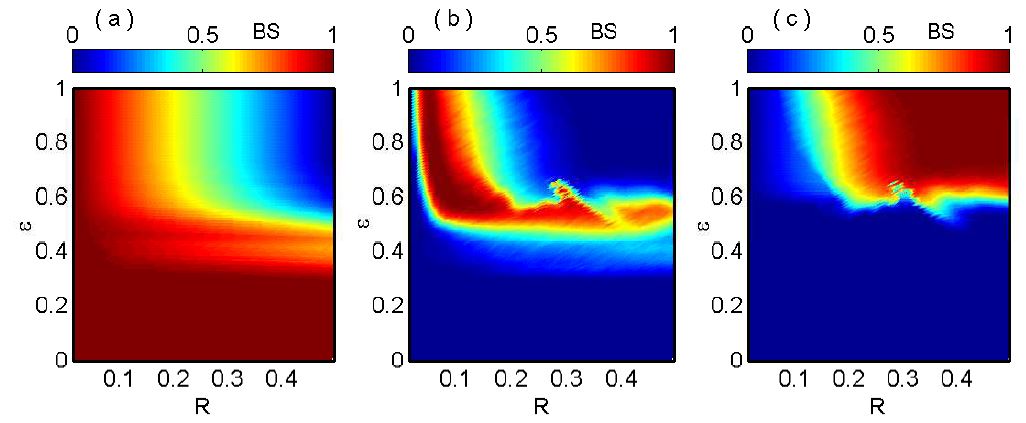}}
 	\caption{ Coexistence of different dynamical states are quantified by a color bar of basin stability measure in $R-\epsilon$ parameter space of nonlocally coupled MG oscillators: (a) incoherent, (b) chimera and (c) coherent states.}
 	\label{fig5}
 \end{figure}

 To explore the complete scenario of variation of BS by simultaneously varying the coupling strength $\epsilon$ and coupling radius $R$, we compute the phase diagram in $R-\epsilon$ plane for the range of $R \in [0, 0.5]$ and $\epsilon \in [0, 1]$.  In the $R-\epsilon$ parameter space of nonlocally coupled MG oscillator, the BS for incoherent, chimera and coherent states are shown in Figs. \ref{fig5}(a), \ref{fig5}(b), and \ref{fig5}(c) respectively. From Fig. \ref{fig5}(a), it is noted that the BS of incoherent state is equal to `1' for lower values of $\epsilon$ and for all the range of $R$. By increasing the value of $\epsilon$ upto a certain value, BS for incoherent state decreases and the probability of acquire more and more points in the basin volume for chimera and coherent states increases. Further increasing of $R$ and/or $\epsilon$, the BS for the incoherent state gradually decreases and finally at the top right corner of Fig. \ref{fig5}(a), the value of BS for incoherent state becomes `0'. Fig. \ref{fig5}(b) corresponds the probability of occurrence of chimera states develops after $\epsilon\simeq0.3$ and probability becomes high near $\epsilon=0.5$. But further increment of $\epsilon$, this high probability will persists if we decrease the values of $R$. From Fig. \ref{fig5}(c), it is easy to conclude that for lower values of $R$ or $\epsilon$ or both, the occurrence of coherent state is impossible. It's BS rises as $R$ and $\epsilon$ increases from $0.01$ and $0.5$ respectively, finally for enough coupling strength and coupling radius the BS of coherent states becomes `1'. Upon increasing the values of $R$ and $\epsilon$ this BS value of coherent state preserves.

 {\subsection*{Globally Coupled Network}}
 Next to testify the universality of basin stability measure for the quantification of coexistence of different dynamical states, we consider the network of globally coupled Mackey-Glass oscillators with additional intensity-dependent term. The mathematical model equations for globally coupled Mackey-Glass system are given by
 \begin{equation}
 \begin{array}{lcl}
 \dot x_i=-\overline{a}x_i+\frac{bx_i(t-\tau)}{1+x_i^{10}(t-\tau)}+\epsilon(X-x_i), \;\;\;\;\; i=1,...,N,
 \end{array}
 \end{equation}
 where $\overline{a}=a+\alpha(x_i^2+x_i^4)$, $X=\frac{1}{N}\sum\limits_{i=1}^{N}x_i$, $\epsilon$ is the coupling strength and $\alpha$ is the intensity-dependent parameter.  Without coupling strength (i.e. $\epsilon=0.0$) and intensity-dependent parameter (i.e. $\alpha=0.0$), individual oscillators oscillate chaotically for the set of system parameters $a=1, b=2$ and $\tau=2$. The dynamics of individual oscillator remains qualitatively same i.e. in chaotic states for smaller values of $\alpha$ in [0.0, 0.03], becomes periodic for its higher value $0.03 <\alpha<0.25$ and for $\alpha\ge 0.25$ the oscillator becomes unbounded. We fix $\alpha=0.002$, so that individual oscillator is in chaotic state. By introducing this intensity parameter $\alpha$, the system becomes more multistable by increasing the number of fixed points. Finally it increases the number of multistable attractors and the coexistence of different collective states emerges depending upon the initial conditions. In the absence of intensity parameter $\alpha$ i.e. $\alpha=0.0,$ the globally coupled network (2) is either incoherent or fully synchronize states depending on the value of coupling strength $\epsilon$.  Here the intensity parameter $\alpha$ plays a crucial role for the emergence of chimera states in globally coupled network.
 
\begin{figure}[ht]
		\centerline
		{	\includegraphics[scale=0.4]{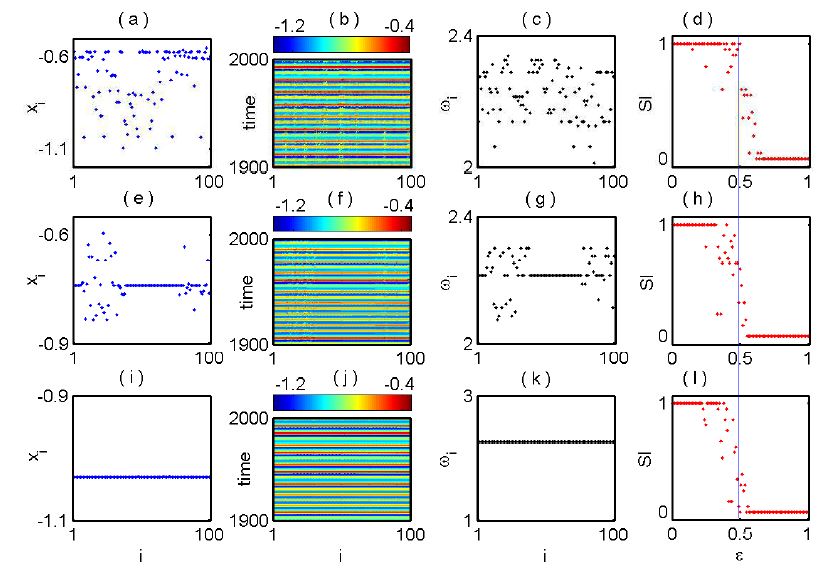}}
	\caption{ Globally coupled Mackey-Glass oscillators: The coexistence of different collective states at a fixed value of coupling strength $\epsilon=0.5$. First, second, third and fourth panels show the snapshots of $x_i$ (blue points), spatiotemporal plot, time average phase velocity (black points) and variation of SI (red points) by changing coupling strength $\epsilon$ respectively. Again first, second and third rows are the results for different initial conditions $a_{ij}=(-1)^j\frac{i-j}{2027}, a_{ij}=(-1)^j\frac{i+j}{2001}$, and $a_{ij}=(-1)^j\frac{i+j}{247}$  respectively for $i=1,2,...,N$ and $j=1,2,...,n=25.$ A blue line is marked along the coupling strength $\epsilon=0.5$. Blue line intersect the SI value at `1' in (d) which characterize the incoherent states corresponds to (a) and similarly it cuts the SI values in $(0,1)$ in (h) and `0' in (l) which resembling the chimera and coherent states correspond to the (e) and (i) respectively. Here $N=100$.} 	
	\label{fig6}
\end{figure}

\par In Fig. \ref{fig6}, we demonstrate the emergence of incoherent and coherent dynamics together with chimera states at a fixed value of coupling strength $\epsilon=0.5$ and different values of initial history functions of each oscillators. Such coexistence behaviors are characterized by a statistical measure SI. We define the initial history functions of coupled network (2) as $\phi_i(t)=\sum\limits_{j=0}^{25}a_{ij}\frac{t^j}{j!}$, where $a_{ij} \in \mathbb{R}$  and $i=1,2,...,N$ for $t\in [-\tau, 0]$.  Next we rigorously search the value of $a_{ij}$ for which the coexistence states emerge.  The incoherent state appears for $a_{ij}=(-1)^j\frac{i-j}{2027}$, the snapshot of $x_i$ and corresponding long time spatiotemporal dynamics are shown in Figs. \ref{fig6}(a) and \ref{fig6}(b) respectively. We also confirm this state by calculating the long time average mean phase velocity $\omega_i$ of each oscillators in Fig. \ref{fig6}(c). In this state, the mean phase velocities are randomly scattered in [2.0, 2.35], which confirm the incoherency between the oscillators. For further characterization, using this fixed value of $a_{ij}$, we compute the value of  SI by changing the coupling strength $\epsilon$ in Fig. \ref{fig6}(d) where the blue line along the coupling strength $\epsilon=0.5$ cuts the SI values at `1'. If we slightly change the coefficient $a_{ij}=(-1)^j\frac{i+j}{2001}$ of initial history function of each oscillators and at the same coupling value $\epsilon=0.5$, we find the mixture of coherent and incoherent population which resemble the chimera states delineate in Fig. \ref{fig6}(e) and corresponding spatiotemporal behavior is illustrated in Fig. \ref{fig6}(f). The time average phase velocity at this particular initial condition is plotted in Fig. \ref{fig6}(g) and the value of SI lies in $(0,1)$ which confirm the chimera state (Fig. \ref{fig6}(h)). The mean phase velocities $\omega_i$ for the coherent group in chimera state are lie on a straight line due to the identical synchronization of the coherent population. The identical synchronization between coherent population in chimera state appeared due to the global coupling topology.   With further tiny disturbance of the coefficient in initial history function of each oscillator by changing $a_{ij}=(-1)^j\frac{i+j}{247}$, all nodes of the network are synchronized (due to all-to-all coupling) and follows the smooth profile represents in Fig. \ref{fig6}(i) and this coherent structure evolve with long time depict (Fig. \ref{fig6}(j)). At this point, the value of all time average phase velocities are identical (Fig. \ref{fig6}(f)) and SI=0 in Fig. \ref{fig6}(l).

 \begin{figure}[ht]
 		\centerline
 		{	\includegraphics[scale=0.48]{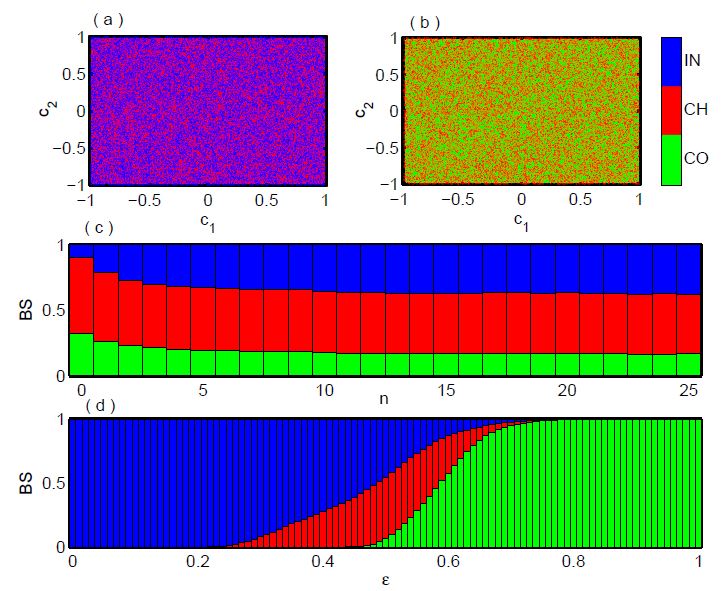}}
 	\caption{ Globally coupled Mackey-Glass systems: Basin of attractions for the coexistence of (a) incoherent and chimera states at $\epsilon=0.34$, (b) chimera and coherent states at $\epsilon=0.56$ where IN, CH and CO stands for incoherent, chimera and coherent states respectively. The basin of stability of incoherent, chimera and coherent states are plotted in (c) with respect to the highest degree $n$ of the polynomial basis represented by blue, red and green colors respectively at the coupling strength $\epsilon=0.57$. (d) The variation of BS for incoherent, chimera and coherent states against the coupling strength $\epsilon$.}
 	\label{fig7}
 \end{figure}

\par  Next we track the basin of attractions of various collective states such as incoherent, chimera and coherent in globally coupled system (2) for fixed value of coupling strength $\epsilon$ and different initial history functions. We  project the infinite dimensional initial history function space of the $i$-th oscillator  to finite dimensional space as $x_i(t)=\phi_i(t)=c_i\sum\limits_{j=0}^{25}(-1)^j\frac{t^j}{j!}$ for $t\in[-\tau, 0]$ where $c_i \in \mathbb{R}$. The coefficients $c_i$ are defined as $c_i=\frac{(i-1)c_1+c_2}{i}, i\geq3$ where $c_1$ and $c_2$ are the random numbers in the range $[-1,1]$. The basin of attractions of different coexistence states are drawn in Figs. \ref{fig7}(a) and \ref{fig7}(b) with respect to $c_1$ and $c_2$  where blue, red and green are associated with incoherent, chimera and coherent dynamical states respectively by taking the fixed coupling strength $\epsilon=0.34$ and $\epsilon=0.56$. In Fig. \ref{fig7}(a) basin of attraction of two states as incoherent and chimera are plotted while Fig. \ref{fig7}(b) represents the basin of attraction of chimera-coherent states. From these two plots, it is clearly enunciated that basin of attraction of the incoherent, chimera and coherent states are highly grappled.

\par The basin stability for chimera state together with coherent and incoherent dynamics in globally coupled network (2) is illustrated in Fig. \ref{fig7}(c, d). The blue, red and green regions are the corresponding incoherent, chimera and coherent states respectively. Figure \ref{fig7}(c) shows the fluctuation of BS of these three collective states tend to slight for sufficiently large $n$ for $\epsilon=0.57$. The associate scenario of the proportion of BS against different coupling strength are portrayed in Fig. \ref{fig7}(d) by taking $n=25$. For small range of $\epsilon$, the incoherent states is dominate and it continues up to $\epsilon=0.2$ and after this value BS of incoherent profile tend to shrinking and chimera states appear with high probability. At $\epsilon$ near equal to $0.47$, the BS of coherent states appear and gradually increases for further increment of $\epsilon$. The three states coexist and it persists up to $\epsilon\simeq0.75$ and more increasing values of $\epsilon$ coherent states are solely dominated with BS unity.
  
\section*{Discussion}
\noindent We have performed the basin stability analysis to quantify different dynamical states, such as incoherent, coherent and chimera states, in nonlocally and globally coupled time-delayed Mackey-Glass oscillators. Usually, the transitions between these states occur as a consequence of parameter changes, such as changes in the coupling strength or the coupling radius. Here, however, we show that various collective states can coexist with different initial states of the oscillators in the network. For this, we project the infinite dimensional initial state space to a finite dimensional space by expanding the initial history function into a polynomial basis. By changing the coefficients of the polynomial basis that defines the initial history function, we have obtained the spatiotemporal dynamics and the time average mean phase velocity of each state. We were thus able to describe transition scenarios among different states with respect to initial conditions, and we were also able to quantify the sensitive dependence on the initial history function. Such collective states have been characterized through a statistical measure SI and also confirmed by plotting the time average mean phase velocity. In amplitude chimera states, amplitudes are drifting whereas frequencies are locked \cite{am_fr_chimera} and corresponding mean phase velocity has the same features (smooth profile) for incoherent, chimera and coherent states. Accordingly, the mean phase velocity can not always distinguish the chimera states from incoherent and coherent states. To overcome these limitations, we have used the SI measure, since different states are characterized by different values of SI. Thus, based on the SI measure, we have developed the basin stability framework and elucidated the variation of the basin stability against the coupling strength. When using the basin stability measure, we can effectively consider a large number of initial condition of each oscillator in a coupled network, and we have shown how the basin stability varies for incoherent, chimera and coherent states in the parameter space of defined by the coupling radius $R$ and the coupling strength $\epsilon$. 

We note that the basin stability approach for chimera states can provide additional information on the initial history function of delayed neuronal models, where multi-stability is common. We also anticipate that our quantification measure of different dynamical states based on the basin stability  can provide a good recipe to obtain stable and robustly synchronized states in complex dynamical systems and neuronal networks.

\section*{Methods}
\noindent {\bf Strength of incoherence Measure.}
To characterize different collective dynamical states, we use a quantitative measure as strength of incoherence recently proposed by Gopal et al. \cite{lakshman_measure}. To calculate this measure, we first introduce difference dynamical variables $w_i$ defined as $w_i=x_{i+1}-x_i$ for $i=1,2,...,N$. Then the standard deviation $\sigma$ is
\begin{equation}
\begin{array}{lcl}
\sigma=\langle \sqrt{\frac{1}{N}\sum\limits_{i=1}^{N}(w_i-\langle w \rangle)^2} \rangle _{t},
\end{array}
\end{equation}
where $\langle w \rangle=\frac{1}{N}\sum\limits_{i=1}^{N}w_i(t)$ and $\langle \cdots \rangle_t$ denotes the time average. This standard deviation measure $\sigma$ distinguish between coherent and incoherent dynamics for taking zero and nonzero values respectively but could not separate chimera states from incoherent states. To overcome this difficulty, we calculate the local standard deviation as $\sigma(m)$. To distinguish chimera and incoherent states, we divide the number of oscillators into $M$ number of bins of equal length $n=N/M$. The local standard deviation $\sigma(m)$ define as
\begin{equation}
\begin{array}{lcl}
\sigma(m)=\langle \sqrt{\frac{1}{n}\sum\limits_{j=n(m-1)+1}^{mn}(w_i-\langle w \rangle)^2} \rangle _{t}\; \mbox{for} \;\; m=1,2,...,M.
\end{array}
\end{equation}
The above quantity $\sigma(m)$ is calculated for every successive $n$ number of oscillators and the strength of incoherence (SI) is defined as
\begin{equation}
\begin{array}{lcl}
SI=1-\frac{\sum\limits_{m=1}^{M}s_m}{M},  \;\;\;\;\;\;  s_m=\Theta[\delta-\sigma(m)],
\end{array}
\end{equation}
where $\Theta(\cdot)$ is the Heaviside step functions and $\delta$ is a predefined threshold which is reasonably small, in general the values of $\delta$ is taken as certain percentage value of the magnitude of the difference between $x_{i,max}$ and $x_{i,min}$. Depending on the values of $\delta$ and $\sigma(m)$, $s_m$ takes `1' or `0' and consequently the values of SI=1 or SI=0 or 0 $<$ SI $<$ 1 represent incoherent, coherent, and chimera states respectively. To calculate SI, we choose an optimal bin size $M=20$ and $\delta=0.05$.
\\
\par \noindent {\bf Mean phase velocity.}
To confirm the existence of different states such as incoherent, coherent and chimera states in a network of coupled oscillators, we compute the time average mean phase velocity $\omega_i$ for each oscillator and it is defined as
\begin{equation}
\begin{array}{lcl}
\omega_i=\frac{2\pi M_i}{\Delta T},  \;\;\;\;\;\;\;\;\;\;\; i=1,2,...,N,
\end{array}
\end{equation}
where $\Delta T$ is the sufficiently large time interval and $M_i$ is the number of maxima of the time series $x_i(t)$ of $i$-th oscillators during the time interval $\Delta T$.\\
\par \noindent {\bf Basin stability measure.}
The space for initial state of time delay differential equation (DDE) is an infinite dimensional function space. Any function from the initial state space can be expanded as a linear combination of an orthogonal or a nonorthogonal basis. For numerical computation, we adopt $P_n$, the space of all polynomials of degree at most $n$ as initial function space of the DDE, which is a $n+1$ dimensional subspace of the polynomial space with a basis $\{1, t, t^2,...,t^n\}$. Other types of basis such as trigonometric, Legendre or Bernstein etc. may be considered as the space. For any $\Phi(t)\in P_n$, there will exists unique $(a_{0},a_{1},a_{2},...,a_{n}) \in \mathbb{R}^{n+1} $ such that $x(t)=\Phi(t)=\sum\limits_{j=0}^{25}a_{j}\frac{t^j}{j!}$ for $t \in [-\tau, 0]$ where $\tau$ is the delay time and $a_{j} \in \mathbb{R}$.
\par In this context, our coupled Mackey-Glass systems can be written as \\
\begin{equation}
\begin{array}{lcl}
\dot x_i=-ax_i+\frac{bx_i(t-\tau)}{1+x_i^{10}(t-\tau)}+K\sum\limits_{j=1}^{N}H_{ij}(x_j-x_i),  i=1,...,N,
\end{array}
\end{equation}
where $H_{ij}$ represents the coupling matrix of the network of coupled dynamical systems and $K$ is the coupling strength. The set denoted by $P_n^N$ for the initial history function of the entire coupled system (7) which is a $N$ times product of $P_n$ space. So the initial history function of the $i^{th}$ oscillator can be written as
$x_i(t)=\Phi_i(t)=\sum\limits_{j=0}^{n}a_{ij}\frac{t^j}{j!}$ for $t\in[-\tau, 0]$, where $a_{ij}\in \mathbb{R},  j=0,1,...,n$ and $i=1,2,...,N.$ Here $n$, the number of the basis functions and the range of $a_{ij}$ play an important role for the calculation of basin stability.
\par For numerical computation, we choose the values of $a_{ij}, (j=1,2,...,n, i=1,2,...,N) $ randomly from $[-1,1]$ which generates the random initial history function for each oscillator. Then we integrate the entire system (7) for sufficiently large number $V$ (say) of initial history function for which $V_s$ number (say) of initial function finally arrives at the corresponding particular state (in our study, the states are incoherent, coherent or chimera). Then the BS of that particular state can be estimated as $\frac{V_s}{V}$. The value of BS lies between 0 and 1. BS=0 means the state is unstable for any random initial conditions and it is monostable for any random perturbation when BS=1. $0<\mbox{BS}<1$ corresponds to the probability of getting the multistable states for any random initial history functions.\\
\par \noindent {\bf Numerical simulations.}
To simulate the coupled dynamical networks (1) and (2), we use the modified Heun method with step size $dt=0.01$. For basin stability measure, we take the sufficiently large number ($T=5000$) of values of the coefficients $a_{ij}$ in the history function of the time-delayed coupled network randomly from $[-1,1]$.  For the simulation of coupled time delay systems, the history function of each oscillator is defined as
\begin{equation}
\begin{array}{lcl}
x_i(t)=\Psi_i(t)=\sum\limits_{j=0}^{n}a_{ij}\frac{t^j}{j!},  \;\;\;\;\;\;\;\; \mbox{for}\;\;\; t\in[-\tau, 0],
\end{array}
\end{equation}
where the values of $a_{ij}, j=0,1,...,n$ and $i=1,2,...,N$ are drawn randomly from $[-1,1]$.\\\\

 \noindent \textbf{Acknowledgments} \\
 This research was supported by the Slovenian Research Agency (Grants P5-0027 and J1-7009). D.G. was supported by SERB-DST (Department of Science and Technology), Government of India (Project no. EMR/2016/001039).
 
 \noindent \\ \textbf{Author contributions} \\
 S.R., B.K.B., M.P., and D.G. designed and performed the research as well as wrote the paper.


\begin{thebibliography}{25}

\bibitem{kuramoto} Kuramoto, Y. $\&$ Battogtokh, D. Coexistence of Coherence and Incoherence in Nonlocally Coupled Phase Oscillators. \textit{Nonlinear Phenom. Complex Syst.} {\bf 5}, 380-385 (2002).
\bibitem{strogatz} Abrams, D. M. $\&$ Strogatz, S. H. Chimera States for Coupled Oscillators. \textit{Phys. Rev. Lett.} {\bf 93}, 174102 (2004).
\bibitem{global1} Yeldesbay, A., Pikovsky, A. $\&$ Rosenblum, M. Chimeralike States in an Ensemble of Globally Coupled Oscillators. \textit{Phys. Rev. Lett.} {\bf 112}, 144103 (2014).
\bibitem{global2} Chandrasekar, V. K., Gopal, R., Venkatesan, A. $\&$ Lakshmanan, M. Mechanism for intensity-induced chimera states in globally coupled oscillators. \textit{Phys. Rev. E} {\bf 90}, 062913 (2014).
\bibitem{global3} Mishra, A., Hens, C., Bose, M., Roy, P. K. $\&$ Dana, S. K. Chimeralike states in a network of oscillators under attractive and repulsive global coupling. \textit{Phys. Rev. E} {\bf 92}, 062920 (2015).
\bibitem{global4} Sethia, G. C. $\&$ Sen, A. Chimera States: The Existence Criteria Revisited. \textit{Phys. Rev. Lett.} {\bf 112}, 144101 (2014).
\bibitem{global5} B\"{o}hm, F., Zakharova, A., Sch\"{o}ll, E. $\&$ L\"{u}dge, K. Amplitude-phase coupling drives chimera states in globally coupled laser networks. \textit{Phys. Rev. E} {\bf 91}, 040901(R) (2015).
\bibitem{global6} Schmidt, L., $\&$ Krischer, K. Clustering as a Prerequisite for Chimera States in Globally Coupled Systems. \textit{Phys. Rev. Lett.} {\bf 114}, 034101 (2015).
\bibitem{global7} Schmidt, L., $\&$ Krischer, K. Chimeras in globally coupled oscillatory systems: From ensembles of oscillators to spatially continuous media. \textit{Chaos} {\bf 25}, 064401 (2015).
\bibitem{hr_bera1} Bera, B. K., Ghosh, D. $\&$ Lakshmanan, M. Chimera states in bursting neurons. \textit{Phys. Rev. E} {\bf 93}, 012205 (2016).
\bibitem{laing} Laing, C. R. Chimeras in networks with purely local coupling. \textit{Phys. Rev. E} {\bf 92}, 050904(R) (2015).
\bibitem{hr_bera2} Bera, B. K. $\&$ Ghosh, D. Chimera states in purely local delay-coupled oscillators. \textit{Phys. Rev. E} {\bf 93}, 052223 (2016).
\bibitem{local1} Hizanidis, J., Lazarides, N. $\&$ Tsironis, G. P. Robust chimera states in SQUID metamaterials with local interactions. \textit{Phys. Rev. E} {\bf 94}, 032219 (2014).
\bibitem{hr_bera3} Bera, B. K., Ghosh, D. $\&$ Banerjee, T. Imperfect traveling chimera states induced by local synaptic gradient coupling. \textit{Phys. Rev. E} {\bf 94}, 012215 (2016).
\bibitem{limit} Ulonska, S., Omelchenko, I., Zakharova, A. $\&$ Sch\"{O}ll, E. Chimera states in networks of Van der Pol oscillators with hierarchical connectivities. \textit{Chaos} {\bf 26}, 094825 (2016).	
\bibitem{chaotic} Gu, C., St-Yves, G. $\&$ Davidsen, J. Spiral Wave Chimeras in Complex Oscillatory and Chaotic Systems. \textit{Phys. Rev. Lett.} {\bf 111}, 134101 (2013).
\bibitem{chaotic_map} Omelchenko, I., Maistrenko, Y., H\"{o}vel, P. $\&$ Sch\"{o}ll, E. Loss of Coherence in Dynamical Networks: Spatial Chaos and Chimera States. \textit{Phys. Rev. Lett.} {\bf 106}, 234102 (2011).
\bibitem{lakshman_measure} Gopal, R., Chandrasekar, V. K., Venkatesan, A. $\&$ Lakshmanan, M. Observation and characterization of chimera states in coupled dynamical systems with nonlocal coupling. \textit{Phys. Rev. E} {\bf 89}, 052914 (2014).
\bibitem{hr_ijbc} Hizanidis, J., Kanas, V., Bezerianos, A. $\&$ Bountis, T. Chimera states in networks of nonlocally coupled Hindmarsh–Rose neuron models. \textit{Int. J. Bifurcat. Chaos} {\bf 24}, 1450030 (2014).
\bibitem{chimera_modular} Hizanidis, J., Kouvaris, N. E., Zamora-L\'{o}pez, G., D\'{i}az-Guilera, A. $\&$ Antonopoulos, C. G. Chimera-like States in Modular Neural Networks. \textit{Sci. Rep.} {\bf 6}, 19845 (2016).
\bibitem{multiplex1} Majhi, S., Perc, M. $\&$ Ghosh, D. Chimera states in uncoupled	neurons induced by a multilayer	structure. \textit{Sci. Rep.} {\bf 6}, 39033 (2016).	
\bibitem{chimera_multiplex} Maksimenko, V. A., Makarov, V. V., Bera, B. K., Ghosh, D., Dana, S. K., Goremyko, M. V., Frolov, N. S., Koronovskii, A. A. $\&$ Hramov, A. E. Excitation and suppression of chimera states by multiplexing. \textit{Phys. Rev. E} {\bf 94}, 052205 (2016).
\bibitem{multiplex2} Ghosh, S. $\&$ Jalan, S. Emergence of Chimera in Multiplex Network. \textit{Int. J. Bifur. Chaos} {\bf 26}, 1650120 (2016).
\bibitem{multiplex3} Ghosh, S., Kumar, A., Zakharova, A. $\&$ Jalan, S. Birth and death of chimera: Interplay of delay and multiplexing. \textit{Europhys. Letts.} {\bf 115}, 60005 (2016).
\bibitem{amc} Sethia, G. C., Sen, A. $\&$ Johnston, G. L. Amplitude-mediated chimera states. \textit{Phys. Rev. E} {\bf 88}, 042917 (2013).
\bibitem{gcc} Sheeba, J. H., Chandrasekar, V. K. $\&$ Lakshmanan, M. Globally clustered chimera states in delay-coupled populations. \textit{Phys. Rev. E} {\bf 79}, 055203(R) (2009).
\bibitem{cd_prl} Zakharova, A., Kapeller, M. $\&$ Sch\"{o}ll, E. Chimera Death: Symmetry Breaking in Dynamical Networks. \textit{Phys. Rev. Lett.} {\bf 112}, 154101 (2014).
\bibitem{breath1} Abrams, D. M., Mirollo, R., Strogatz, S. H. $\&$ Wiley, D. A. Solvable Model for Chimera States of Coupled Oscillators. \textit{Phys. Rev. Lett.} {\bf 101}, 084103 (2008).
\bibitem{imperfect_chi} Kapitaniak, T., Kuzma, P., Wojewoda, J., Czolczynski, K. $\&$ Maistrenko, Y. Imperfect chimera states for coupled pendula. \textit{Sci. Rep.} {\bf 4}, 6379 (2014).
\bibitem{travelling_chi} Xie, J., Knobloch, E. $\&$ Kao, H. C. Multicluster and traveling chimera states in nonlocal phase-coupled oscillators. \textit{Phys. Rev. E} {\bf 90}, 022919 (2014).	
\bibitem{spiral_chi} Li, B. W. $\&$ Dierckx, H. Spiral wave chimeras in locally coupled oscillator systems. \textit{Phys. Rev. E} {\bf 93}, 020202(R) (2016).
\bibitem{opto-elect} Hagerstrom, A., Murphy, T. E., Roy, R., H\"{o}vel, P., Omelchenko, I. $\&$ Sch\"{o}ll, E. Experimental observation of chimeras in coupled-map lattices. \textit{Nat. Phys.} {\bf 8}, 658-661 (2012).
\bibitem{chemical_exp} Tinsley, M. R., Nkomo, S. $\&$ Showalter, K. Chimera and phase-cluster states in populations of coupled chemical oscillators. \textit{Nat. Phys.} {\bf 8}, 662-665 (2012).
\bibitem{chemical_exp2} Nkomo, S., Tinsley, M. R. $\&$ Showalter, K. Chimera States in Populations of Nonlocally Coupled Chemical Oscillators. \textit{Phys. Rev. Lett.} {\bf 110}, 244102 (2012).
\bibitem{electronic2} Larger, L., Penkovsky, B. $\&$ Maistrenko, Y. Virtual Chimera States for Delayed-Feedback Systems. \textit{Phys. Rev. Lett.} {\bf 111}, 054103 (2013).
\bibitem{electronic} Gambuzza, L. V., Buscarino, A., Chessari, S., Fortuna, L., Meucci, R. $\&$ Frasca, M. Experimental investigation of chimera states with quiescent and synchronous domains in coupled electronic oscillators. \textit{Phys. Rev. E} {\bf 90}, 032905 (2014).
\bibitem{electro} Wickramasinghe, M. $\&$ Kiss, I. Z. Spatially Organized Dynamical States in Chemical Oscillator Networks: Synchronization, Dynamical Differentiation, and Chimera Patterns. \textit{PLoS ONE} {\bf 8}, e80586 (2013).
\bibitem{electro2} Schmidt, L., Sch\"{o}nleber, K., Krischer, K. $\&$ Vladimir Garc\'{i}a-Morales, V. Coexistence of synchrony and incoherence in oscillatory media under nonlinear global coupling. \textit{Chaos} {\bf 24}, 013102 (2014).
\bibitem{opto_electronic} Larger, L., Penkovsky, B. $\&$ Maistrenko, Y. Laser chimeras as a paradigm for multistable patterns in complex systems. \textit{Nat. Commun.} {\bf 6}, 7752 (2015).
\bibitem{boolean} Rosin, D. P., Rontani, D. $\&$ Gauthier, D. J. Synchronization of coupled Boolean phase oscillators. \textit{Phys. Rev. E} {\bf 89}, 042907 (2014).
\bibitem{opt_comb} Viktorov, E. A., Habruseva, T., Hegarty, S. P., Huyet, G. $\&$ Kelleher, B. Coherence and Incoherence in an Optical Comb. \textit{Phys. Rev. Lett.} {\bf 112}, 224101 (2014).
\bibitem{mechanical} Martens, E. A., Thutupalli, S., Fourriere, A. $\&$ Hallatschek, O. Chimera states in mechanical oscillator networks. \textit{Proc. Nat. Acad. Sci. USA} {\bf 110}, 10563-10567 (2013).
\bibitem{chimera_rev} Panaggio, M. J. $\&$ Abrams, D. M. Chimera states: coexistence of coherence and incoherence in networks of coupled oscillators.  \textit{Nonlinearity} {\bf 28}, R67 (2015).
\bibitem{power1} Motter, A. E., Myers, S. A., Anghel, M. $\&$ Nishikawa, T. Spontaneous synchrony in power-grid networks. \textit{Nat. Phys.} {\bf 9(3)}, 191-197 (2013).
\bibitem{power2} D\"{o}rfler, F., Chertkov, M. $\&$ Bullo, F. Synchronization in complex oscillator networks and smart grids. \textit{Proc. Nat. Acad. Sci. USA} {\bf 110(6)}, 2005-2010 (2013).
\bibitem{social} Gonz\'{a}lez-Avella, J. C., Cosenza, M. G., $\&$ Miguel, M. S. Localized coherence in two interacting populations of social agents. \textit{Physica A.} {\bf 399}, 24-30 (2014).
\bibitem{uhsws1} Rattenborg, N. C., Amlaner, C. J. $\&$ Lima, S. L. Behavioral, neurophysiological and evolutionary perspectives on unihemispheric sleep. \textit{Neurosci. Biobehav. Rev.} {\bf 24}, 817-842 (2000).
\bibitem{uhsws2} Rattenborg, N. C. Do birds sleep in flight? \textit{Naturwissenschaften} {\bf 93}, 413-425 (2006).
\bibitem{ba_njp} Martens, E. A., Panaggio, M. J. $\&$ Abrams, D. M. Basins of attraction for chimera states. \textit{New J. Phys.} {\bf 18}, 022002 (2016).
\bibitem{random} Larger, L., Penkovsky, B. $\&$ Maistrenko, Y. Virtual Chimera States for Delayed-Feedback Systems. \textit{Phys. Rev. Lett.} {\bf 111}, 054103 (2013).
\bibitem{random2} Nkomo, S.,  Tinsley, M. R. $\&$ Showalter, K. Chimera States in Populations of Nonlocally Coupled Chemical Oscillators. \textit{Phys. Rev. Lett.} {\bf 110}, 244102 (2013).	 

\bibitem{bs_nature} Menck, P. J., Heitzig, J.,	Marwan, N. $\&$ Kurths, J. How basin stability complements the linear-stability paradigm. \textit{Nat. Phys.} {\bf 9}, 89-92 (2013).
\bibitem{our2017} Rakshit, S., Bera, B. K., Majhi, S., Hens, C. $\&$ Ghosh, D. Basin stability measure of different	steady states in coupled oscillators. \textit{Sci. Rep.} {\bf 7}, 45909 (2017).
\bibitem{bs_timedelay} Leng, S., Lin, W. $\&$ Kurths, Y. Basin stability in delayed dynamics. \textit{Sci. Rep.} {\bf 6}, 21449 (2016).

\bibitem{pwr_grd1} Machowski, J., Bialek, J. W. $\&$ Bumby, J. R. Power System Dynamics: Stability and Control (Wiley, 2008).
\bibitem{pwr_grd2} Menck, P. J. $\&$ Kurths, J. Topological identification of weak points in power grids. In \textit{Nonlinear Dynamics of Electronic Systems, Proceedings of NDES 2012,} 1-4 (VDE,2012).
\bibitem{njp} Schultz, P., Heitzig, J. $\&$ Kurths. J. Detours around basin stability in power networks. \textit{New J. Phys.} {\bf 16}, 125001 (2014).
\bibitem{epjst} Ji, P. $\&$ Kurths, J. Basin stability of Kuramoto-like model in small networks. \textit{The European Physical Journal Special Topics} {\bf 12}, 2483-2491 (2014).
\bibitem{bspre} Maslennikov, O. V.,  Nekorkin, V. I. $\&$ Kurths, J. Basin stability for burst synchronization in small-world networks of chaotic slow-fast oscillators. \textit{Phys. Rev. E} {\bf 92}, 042803 (2015).	
\bibitem{mg} Mackey, M. C. $\&$ Glass, L. Oscillation and chaos in physiological control systems. \textit{Science} {\bf 197}, 287-289 (1977).	
\bibitem{rev}  Sevilla-Escoboza, R.,  Buld\'{u}, J. M.,  Pisarchik, A. N.,  Boccaletti, S. $\&$  Guti\'{e}rrez, R. Synchronization of intermittent behavior in ensembles of multistable dynamical systems. \textit{Phys. Rev. E} {\bf 91}, 032902 (2015).
\bibitem{am_fr_chimera} Gopal, R., Chandrasekar, V. K., Senthilkumar, D. V., Venkatesan, A. $\&$ Lakshmanan, M. Effect of asymmetry parameter on the dynamical states of nonlocally coupled nonlinear oscillators. \textit{Phys. Rev. E} {\bf 91}, 062916 (2015).



\end{thebibliography}
\end{document}